\documentclass[aps,prl, groupedaddress,showpacs,showkeys]{revtex4}
\usepackage{amssymb,bm}
\usepackage{graphicx}
\usepackage{amsmath}
\allowdisplaybreaks

\begin{document}

\title{Coulomb corrections to electron scattering on the extended source and the proton charge radius.}

\author{R.N. Lee}\email{R.N.Lee@inp.nsk.su}
\author{A. I. Milstein}\email{A.I.Milstein@inp.nsk.su}
\affiliation{G.I.Budker Institute of Nuclear Physics, 630090 Novosibirsk, Russia}

\date{\today}

\begin{abstract}
It is shown that the account for the proton charge form factor in the Coulomb corrections to the electron-proton scattering cross section noticeably diminishes the difference between the value of  the proton charge radius $r_{\mathrm{E}}$, extracted from the  $ep$ scattering data, and that following  from the   muonic hydrogen data. For the electron energy much higher than the electron mass but much smaller than $r_{\mathrm{E}}^{-1}\approx 230$~MeV, the relative correction has the universal form $\delta r_{\mathrm{E}}/r_{\mathrm{E}}=-\pi\alpha/2$, where $\alpha$ is the fine structure constant.
\end{abstract}



\maketitle

\section{Introduction}
One of the unsolved puzzles in the elementary particle physics widely discussed now is the huge difference, five standard deviations, between the value of  the proton charge radius  $r_{\mathrm{E}}$ extracted from the $ep$ scattering data, Refs.\cite{Bernauer2010,Bernauer2013},  and that following  from the muonic hydrogen data, Refs.~\cite{Pohl2010,Anto2013}. Though  careful revisions of the theoretical results for the muonic hydrogen atom have been performed, no contribution which could  noticeably shift the value of the proton  charge radius was found, see review in Ref.~\cite{Jent2011}. Concerning the result following from the $ep$ scattering data, the most  problematic point is the account for the radiative corrections, and consensus  was not achieved whether this account was performed correctly, see discussion in Refs.~ \cite{Arrington2011,Bernauer2011}. Though there are many papers devoted to the calculation of the radiative corrections to  $ep$ scattering, see, e.g., Refs.~\cite{Lewis1956,Tsai1961,Bottino1971,Rosen2000,Arrington2007,Borisyuk2007}, the authors of the experimental paper \cite{Bernauer2010} and the author of Ref.~\cite{Arrington2011} came to the opposite conclusions on the importance of account for the proton form factor in the calculations of the Coulomb corrections to the $ep$ scattering cross section, see also Refs. \cite{Bernauer2011,Bernauer2013}.

In the present paper we derive explicitly the shift  $\delta r_{\mathrm{E}}$ of the proton charge radius which appears due to the account for the proton charge form factor in the leading Coulomb corrections. These corrections are the contribution of the two-photon exchange amplitude calculated in the external field approximation. The derived shift is essentially larger than the model uncertainties declared by the authors of Ref. \cite{Bernauer2013}. Moreover, it diminishes noticeably the difference between the value of the proton charge radius following from the  muonic hydrogen data and that following from the $ep$ scattering data.

\section{Coulomb corrections}

In the $ep$ scattering experiment, the most accurate value of the proton charge radius are obtained at the smallest value  of the momentum transfer $\bm Q=\bm p_2-\bm p_1$, where $\bm p_1$ and $\bm p_2$ are the initial and final electron momenta, respectively, we set $\hbar=c=1$ throughout the paper. In the experiment \cite{Bernauer2010}, the minimal value of $Q$ was $60$~MeV which is much smaller than the proton mass. Besides, the lowest electron energy $ E =180$~MeV in this experiment is also essentially smaller than the proton mass, so that the external field approximation is appropriate to study the influence of the charge form factor on the Coulomb corrections to the $ep$ scattering amplitude and the cross section.

Let us consider the electron scattering cross section in the external potential $V(r)$, having the Fourier transform $V_F(Q^2)=-4\pi\alpha  F(Q^2)/Q^2$, where ${F}(Q^2)$ is the charge form factor of the proton, and $\alpha\approx 1/137$ is the fine structure constant. The cross section  reads (see, e.g., Ref.~\cite{BLP82} )
\begin{equation}\label{cs}
\frac{d\sigma}{d\Omega}= \sum_{\lambda_2} |M_{\lambda_2\lambda_1}|^2\,,\quad  M_{\lambda_2\lambda_1}=-\frac{ E }{2\pi}{u}^\dagger_{\lambda_2\bm p_2}
\int d\bm r \exp{(-i\bm p_2\cdot\bm r)}V(r)\psi_{\lambda_1\bm p_1}^{(in)}(\bm r) \,,
\end{equation}
where $ E =\sqrt{ p_{1,2}^2+m^2}$ is the electron energy, $m$ is the electron mass, $ \psi_{\lambda_1\bm p_1}^{(in)}(\bm r )$ is a positive-energy  solution  of the Dirac equation in the external field, the asymptotic form of  $ \psi_{\lambda_1\bm p_1}^{(in)}(\bm r )$ at large $\bm r$ contains the plane wave and the spherical divergent wave, $ u_{\lambda_2\bm p_2}$ is the free Dirac spinor normalized to $ u_{\lambda_2\bm p_2}^\dagger u_{\lambda_2\bm p_2}=1$, $\lambda_1=\pm 1$ and $\lambda_2=\pm 1$ enumerate the independent solutions of the Dirac equation. It is shown in Ref.~\cite{DLMR2013} that
  the wave function  $ \psi_{\lambda_1\bm p_1}^{(in)}(\bm r )$ has the form
\begin{align}\label{wf}
&\psi_{\lambda_1{\bm p_1}}^{(in)}({\bm r})=\left[g_0(\bm p_1,\bm r)+\bm\alpha\cdot\bm g_1(\bm p_1,\bm r)
+\bm\Sigma\cdot\bm g_2(\bm p_1,\bm r)\right]u_{\lambda_1{\bm p_1}} \,,
\end{align}
where $\bm\alpha=\gamma^0\bm\gamma$, $\bm\Sigma=-\gamma^5\gamma^0\bm\gamma$, and $\gamma^\mu$ are the Dirac matrices.  Note that the perturbation expansion of the functions $g_0(\bm p_1,\bm r)$,
$\bm g_1(\bm p_1,\bm r)$, and $\bm g_2(\bm p_1,\bm r)$  starts from the terms $V^0$, $V^1$ and $V^2$, respectively. We are going to calculate the matrix element  $M_{\lambda_2\lambda_1}$ taking into account the contributions  $V^1$ and $V^2$ only (the one-photon exchange and the two-photon exchange). Therefore, we can neglect the term with $\bm g_2(\bm p_1,\bm r)$.
Let us introduce the quantities
\begin{eqnarray}\label{GG}
&&(G_0,\,\bm G_{1})=-\frac{ E }{2\pi}\int \!\!d\bm r\,\exp{(-i\bm p_2\cdot\bm r )}V(r)(g_0(\bm p_1,\bm r),\, \bm g_1(\bm p_1,\bm r))\,.
\end{eqnarray}
Due to the parity conservation, the function $\bm G_{1}$ can be represented as $f_1\bm p_1+f_2\bm p_2$, where $f_{1,2}$ depend on $\bm p_1\cdot\bm p_2$. Therefore, using the Dirac equation for $u_{\lambda_1{\bm p_1}}, u^{\dagger}_{\lambda_2{\bm p_2}}$ and neglecting all terms of order $m/ E $ we obtain
\begin{eqnarray}\label{M12}
&& M_{\lambda_2\lambda_1}=T{u}^\dagger_{\lambda_2\bm p_2}{u}_{\lambda_1\bm p_1}\,, \quad T=G_0+\frac{(\bm\nu\cdot\bm G_1)}{\nu^2}\,,\nonumber\\
&& \bm\nu=\frac{(\bm p_1+\bm p_2)}{2 E }\,.
\end{eqnarray}
Remarkably, in Eq.\eqref{M12} the Coulomb corrections enter the scattering amplitude via the overall scalar factor $T$.
Using the conventional perturbation theory we find for the linear and quadratic in $V$ terms of $T=T_1+T_2$,
\begin{eqnarray}\label{T12}
&&T_1=-\frac{ E  V_F(Q^2)}{2\pi}\,,\nonumber\\
&&T_2=\frac{ E }{(2\pi)^4}\int d\bm s\frac{V_F(\chi_+)V_F(\chi_-)[2 E +(\bm\nu\cdot\bm s)/\nu^2]}{s^2+2 E  (\bm\nu\cdot\bm s)-Q^2/4-i0}\,,\nonumber\\
&&\chi_\pm=(\bm s\pm \bm Q/2)^2\,.
\end{eqnarray}
Up to $V^3$ terms, the cross section reads
\begin{equation}
d\sigma=(1+\delta)d\sigma_B, \quad \delta=2\mbox{Re}(T_1^*T_2)/|T_1|^2\,,\label{eq:delta_sigma}
\end{equation}
where $d\sigma_B$ is the Born cross section. Since $T_1$ is the real quantity, one should calculate the real part of $T_2$. Eq.\eqref{eq:delta_sigma} is in agreement with the corresponding result of Ref. \cite{Lewis1956}.

For a pointlike particle, when $F(Q^2)=1$, the Coulomb correction $\delta$ is reduced to the  Feshbach correction $\delta_F$, Ref. \cite{Tsai1961,Fesh1948}
\begin{eqnarray}\label{df}
\delta_F=\frac{\pi\alpha\sin\frac{\theta}{2}}{1+\sin\frac{\theta}{2}}\,,\quad \sin\frac{\theta}{2}=\frac{Q}{2p}\,.
\end{eqnarray}
It was explicitly stated in Ref. \cite{Bernauer2013} that the Coulomb corrections to the cross section were taken into account in the data analysis in the experiment \cite{Bernauer2010} solely via the correction-factor $(1+\delta_F)$. In the next Section we show that this is not sufficient for the accuracy declared in Refs. \cite{Bernauer2010,Bernauer2013}.

\section{Charge radius}
Let us discuss the effect of finite proton charge radius on the Coulomb corrections. We consider small-angle scattering when $Q\ll  E ,\,r_{\mathrm{E}}^{-1}$,
 where $r_{\mathrm{E}}^{-1}\approx 230$~MeV.  In this case $F(Q^2)\approx1-r^2_{\mathrm{E}}Q^2/6$, so that
\begin{eqnarray}\label{G00}
T_1=\frac{2 E \alpha}{Q^2}F(Q^2)\approx 2 E \alpha\left(\frac{1}{Q^2}-\frac{r^2_{\mathrm{E}}}{6}\right)\,.
\end{eqnarray}
Therefore, the  $Q$-independent term in $T_2$ can imitate the effect of $r_{\mathrm{E}}$ in the Born term. Thus, the correction $\delta r_{\mathrm{E}}^2$ coming  from the account of the finite proton charge radius has the form
\begin{eqnarray}\label{dr2}
\delta r_{\mathrm{E}}^2=-\frac{3\alpha}{\pi^2}\, \mbox{Re} \int d\bm s\,\frac{[1-F^2(s^2)][2 E +(\bm\nu\cdot\bm s)]}{s^4[s^2+2 E  (\bm\nu\cdot\bm s)-i0]}\,.
\end{eqnarray}
If $ E \ll r_{\mathrm{E}}^{-1}$, then the main contribution to the integral in Eq. \eqref{dr2} is given by the region $s\sim  E $ and we can replace $1-F^2(s^2)$ by $s^2r^2_{\mathrm{E}}/3$. Then we obtain
\begin{equation}\label{dr2main}
\delta r_{\mathrm{E}}^2=-\pi\alpha r_{\mathrm{E}}^2\,, \quad \frac{\delta r_{\mathrm{E}}}{r_{\mathrm{E}}}=-\frac{\pi\alpha}{2}\approx -0.0115\,.
\end{equation}
Thus, at  $ E \ll r_{\mathrm{E}}^{-1}$ the relative correction  $\delta r_{\mathrm{E}}/r_{\mathrm{E}}$ is independent of the shape of the form factor. It is two times larger than the statistical error and five times larger than the model error presented in Ref. \cite{Bernauer2010}.

For  $ E \gg r_{\mathrm{E}}^{-1}$, the main contribution to the integral in Eq.(\ref{dr2}) is given by the region $s\sim r_{\mathrm{E}}^{-1}$, and we find in this limit,
\begin{equation}\label{dr2eps}
\delta r_{\mathrm{E}}^2=-\frac{12\alpha}{\pi E }\, \int \frac{ds}{s^2}[1-F^2(s^2)]\,.
\end{equation}
In contrast to the asymptotics \eqref{dr2main}, the asymptotics \eqref{dr2eps} is model-dependent. The widely used fit of the form factor has the dipole form
\begin{equation}\label{dipole}
F(Q^2)=\left[1+\frac{Q^2}{\Lambda^2}\right]^{-2}\,,\quad \Lambda=840 \mbox{ MeV}\,.
\end{equation}
For this form factor, the correction Eq.~(\ref{dr2eps}) reads
\begin{equation}\label{dr2epsd}
 \frac{\delta r_{\mathrm{E}}}{r_{\mathrm{E}}}=-\frac{35\alpha\Lambda}{64 E }\,.
\end{equation}

\begin{figure}
\centering
\includegraphics[width=0.7\linewidth]{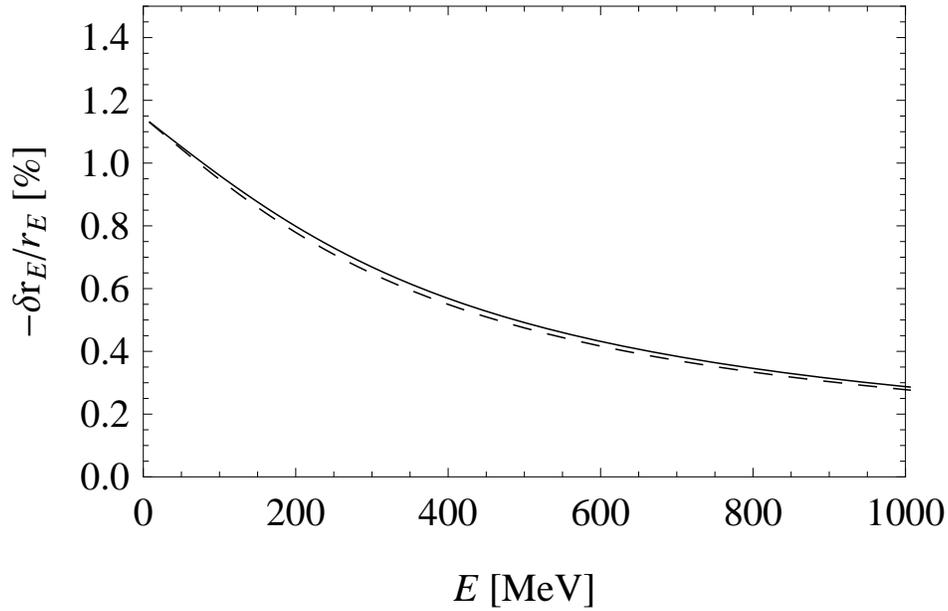}
\caption{The dependence of $\delta r_{\mathrm{E}}/r_{\mathrm{E}}$ on $ E $. Solid line corresponds to the dipole form factor \eqref{dipole}, dashed line corresponds to the form factor in Eq. \eqref{monopole}.}
\label{fig1}
\end{figure}

In Fig.~\ref{fig1} we show the dependence of the relative correction $\delta r_{\mathrm{E}}/r_{\mathrm{E}}$ on $ E $ for the  dipole form factor (solid line). In fact, this dependence is not very sensitive to the shape of the form factor at a given $r_{\mathrm{E}}$. This statement is demonstrated in Fig.~\ref{fig1} by the dashed line obtained for the form factor
\begin{equation}\label{monopole}
F(Q^2)=\left[1+\frac{2Q^2}{\Lambda^2}\right]^{-1}\,,
\end{equation}
having the same expansion  at small $Q^2$ as the expansion of the dipole form factor (\ref{dipole}). One can see that $\delta r_{\mathrm{E}}/r_{\mathrm{E}}$ is very sensitive to the energy $ E $, and becomes $\delta r_{\mathrm{E}}/r_{\mathrm{E}}=-0.82\cdot 10^{-2}$ for $ E =180$~MeV,  corresponding to the minimal  energy in the experiment Ref.\cite{Bernauer2010}. This correction is still essentially (more than three times) larger than the model error presented in Ref.\cite{Bernauer2013}.

\section{Conclusion}

We have demonstrated that, for the model uncertainty declared in Refs.\cite{Bernauer2010,Bernauer2013}, the account for the Coulomb corrections via the factor $(1+\delta_F)$, with $\delta_F$ being the Feshbach correction, Eq. \eqref{df}, is not sufficient.
The account for the proton charge form factor in the Coulomb corrections is essentially larger than the model uncertainty of Refs.\cite{Bernauer2010,Bernauer2013} and diminishes noticeably the difference between the value of  the proton charge radius $r_{\mathrm{E}}$, extracted from the  $ep$ scattering data, and that following  from the muonic hydrogen data. For the electron energy much higher than the electron mass but much smaller than $r_{\mathrm{E}}^{-1}\approx 230$~MeV, the relative correction has the universal form $\delta r_{\mathrm{E}}/r_{\mathrm{E}}=-\pi\alpha/2$.

Certainly, for the energies, corresponding to the experimental conditions in \cite{Bernauer2010}, there are some corrections to Eq. \eqref{dr2} coming from the recoil and virtuality effects. However, they can not change the conclusion on the importance of the account for the proton form factor at the calculation of the Coulomb corrections to provide the accuracy declared in Refs.\cite{Bernauer2010,Bernauer2013}.

\paragraph{Acknowledgments.} The work  has been supported in part  by the Ministry of Education and Science of the Russian Federation and the RFBR grant no. 14-02-00016.

\end{document}